# Electric-field modulation of exchange stiffness in MgO/CoFeB with perpendicular anisotropy

T. Dohi,[1] S. Kanai,[1,2,a)] A. Okada,[1] F. Matsukura,[1,2,3] and H. Ohno[1,2,3]

[1]*Laboratory for Nanoelectronics and Spintronics, Research Institute of Electrical Communication, Tohoku University, 2-1-1 Katahira, Aoba-ku, Sendai 980-8577, Japan*

[2]*Center for Spintronics Integrated Systems, Tohoku University, 2-1-1 Katahira, Aoba-ku, Sendai 980-8577, Japan*

[3]*WPI-Advanced Institute for Materials Research, Tohoku University, 2-1-1 Katahira, Aoba-ku, Sendai 980-8577, Japan*

We observe magnetic domain structures of MgO/CoFeB with a perpendicular magnetic easy axis under an electric field. The domain structure shows a maze pattern with electric-field dependent isotropic period. We find that the electric-field modulation of the period is explained by considering the electric-field modulation of the exchange stiffness constant in addition to the known magnetic anisotropy modulation.

[a)]Corresponding Author: Shun Kanai

Laboratory for Nanoelectronics and Spintronics, Research Institute of Electrical Communication,

Tohoku University, Katahira 2-1-1, Aoba-ku, Sendai 980-8577, Japan

Phone: +81-22-217-5556

Fax: +81-22-217-5555

E-mail address: sct273@riec.tohoku.ac.jp

Capability of mutual control between magnetic and electrical properties in magnetic materials is essential for the study on spintronics. So far, it was reported that by applying an external electric field $E$, one can modulate various magnetic parameters such as the Curie temperature,[1,2] coercive force,[3,4] magnetic anisotropy,[5–7] and damping constant in both ferromagnetic semiconductors and metals.[8–10] These magnetic parameters determine the performance of the spintronics devices. For magnetic tunnel junctions (MTJs), for instance, the thermal stability factor depends on the magnitude of magnetic anisotropy, and the critical current for magnetization switching on the anisotropy and damping constant.[11–13] A switching scheme of a free layer in MTJs as well as the control of domain wall motion and pinning have been shown possible via application of electric fields.[14–20] Although a number of parameters that characterize a ferromagnet are shown to be modulated by electric field as exemplified in these cases, no information is so far available on electric field effect on exchange stiffness, one of the most fundamental constants that characterize a ferromagnetic material. Here we address this question by analyzing domain structures under electric field in a CoFeB/MgO structure with a perpendicular magnetic easy axis. Domain structures of ferromagnetic thin films are determined by a combination of the thickness of the film as well as magnetic parameters such as spontaneous magnetization, magnetic anisotropy, and exchange stiffness constant.[21–26] Therefore, the observation of the domain structures under electric fields is expected to provide the information of the electric-field modulations of these parameters, including the exchange stiffness.

To fabricate a capacitor structures, we deposit Ru (5 nm) and Ta (5 nm) layers, on a thermally oxidized Si substrate, and pattern the layers into a circular bottom electrode with an electrical pad. We then form a gate insulator $Al_2O_3$ (58.8 nm)/MgO (2 nm) and a top metal electrode $Co_{0.2}Fe_{0.6}B_{0.2}$ by lift-off process. All the layers are deposited by dc/rf sputtering, except for the $Al_2O_3$ layer which is formed by atomic layer deposition. The capacitor has a 1 mm diameter, and is annealed at 200°C for 1 h in vacuum under a perpendicular magnetic field of 1 T. The value of the spontaneous magnetization $M_S$ of the CoFeB is determined to be 1.50 T from magnetization measurements. All the measurements in this work are conducted at room temperature. We apply voltage to the device up to ±10 V, which corresponds to an electric field of

±0.11 V/nm in MgO,[7] and the positive voltage corresponds to the top electrode positive with respect to the bottom one.

We measure ferromagnetic resonance (FMR) spectra to determine the effective perpendicular magnetic field and its electric-field modulation.[8] Figure 1 shows the magnetic-field angle $\theta_H$ dependence of resonance fields $H_R$, where $\theta_H$ is measured from the device normal. The lowest $H_R$ at $\theta_H = 0°$ indicates that the CoFeB layer has a perpendicular magnetic easy axis due to the interfacial magnetic anisotropy at MgO/CoFeB.[7,11] The $H_R$ at $\theta_H = 0°$ (90°) decreases (increases) under a positive (negative) electric field, which indicates that the perpendicular anisotropy increases under positive electric fields. By fitting the resonance condition to the dependence as shown by solid lines,[8] we determine the effective magnetic anisotropy energy density $K^{\rm eff}$ at $E = 0$ to be $6.3 \times 10^4$ J/m$^3$, and its areal modulation per electric field to be $dK^{\rm eff}t/dE = 54.6$ fJ/Vm, where $t = 1.5$ nm is the thickness of the CoFeB layer. These values are in line with previous reports.[7,8]

Next, we observe magnetic domain structures at demagnetized state by a polar magneto-optical Kerr effect (MOKE) microscope.[27,28] We demagnetize the CoFeB by applying an alternative perpendicular magnetic field with exponentially decaying amplitude starting from 20 mT. We record differential image between the demagnetized state and remanent state after the application of dc magnetic field to saturate the magnetization. Images in Figs. 2(a)-(c) show thus obtained domain structures at three electric fields of -0.088, 0, and +0.088 V/nm. The domain structures show a maze pattern, in which white (black) region corresponds to the region with up (down) magnetic moments. One can see clearly that the width of the domains decreases under a negative electric field. We analyze the images using two-dimensional fast Fourier transform (2D-FFT) to determine the domain period, which is defined as the distance between the two neighboring black regions.[27] The processed images are shown in Figs. 2(d)-(f), where white dots indicate the regions with large spatial frequency components. A circular shape of the obtained images indicates that the domain pattern is periodic and spatially isotropic in agreement with the previous observation.[27] Figure 2(g) shows the averaged amplitude of the Fourier component obtained by averaging the FFT over the circles. We take the inverse of the peak wavelength as the characteristic domain period $D_P$. Figure 3 summarizes the electric-field $E$ dependence of $D_P$, in which error bars are obtained from five measurements on the arbitrary selected different areas in the device. The $D_P$ shows almost linear dependence on $E$, and $D_P = 1.72$ μm at $E = 0$ changes by ±0.25 μm by applying $E$ of ±0.11 V/nm.

According to the model that describes domain structures in thin magnetic films,[26] $D_P$ is expressed as,

$$D_P = 2(A_S/K^{eff})^{0.5} \exp[4\pi\mu_0(A_S K^{eff})^{0.5}/(M_S^2 t)], \qquad (1)$$

in which $A_S$ is the exchange stiffness constant, and $\mu_0$ is permeability of vacuum. By substituting the experimentally obtained values of $M_S$ and $K^{eff}$, we determine the value of $A_S$ at $E = 0$ to be 8.77±0.15 pJ/m. If we assume the applied electric field modulates only $K^{eff}$ (electric-field independent $M_S$ and $A_S$), we obtain a solid line in Fig. 3 from Eq. (1). While the result reproduces the general trend of the experiment, there is a difference beyond experimental error between this solid line and a linear fit to the experimental points shown as a dotted line in Fig. 3.

There are three possible origins for this difference; (1) the magnitude of $dK^{eff}t/dE$, (2) electric field dependence of $M_S$, and (3) electric field dependence of $A_S$. As to the $dK^{eff}t/dE$, one needs to take the value of $dK^{eff}t/dE$ of ~82 fJ/Vm to reproduce the linear fit, 50% larger than the experimental value of 54.6 fJ/m, which is beyond the range of the experimental error in the determination of $dK^{eff}/dE$. We next consider the change of $M_S$ on $E$. The change of $M_S$ by $\mp 15$ mT with $E$ of ±0.11 V/nm reproduces the fitted line. The applied electric field alters the electron density at the interface, which in principle can change the $M_S$ following the Slater-Pauling curve. However, because the Thomas-Fermi screening, the applied electric field alters the electron density at the interfacial one-monolayer transition metals. The electric-field range used here changes at most ~0.01 electrons per one transition-metal atom at the interface. The Slater-Pauling curve suggests that such a small change of the electron number result in the change of $M_S$ of 1.2 mT (0.08% of $M_S$), which cannot account for the change of 30 mT (2% of $M_S$) needed to reproduce the dotted fit. The difference between the solid and dotted lines, thus, indicates that there is an electric-filed modulation of $A_S$, which is of the order of 3% change (1.3×10$^{-21}$ J/V) in the range of electric field applied in the present experiment. Note that the $A_S$ is an average over the CoFeB film. While theory has to be developed and compared to the present experiment to establish the electronic origin of this modulation, it is reasonable to expect that the modulation of the interface electron density changes the exchange at the interface. We point out that the interface plays a role in determining $A_S$ of a thin layer of ferromagnet as observed in the case of Co in Co/Ru/Co structures, where $A_S$ of Co depends on the strength of the interlayer coupling between the two Co layers.[29] In addition, theoretical calculation predicted also that $A_S$ of thin magnets is sensitive to the electronic structures at the interface of MgO and ferromagnetic layers.[30]

In summary, we have investigated the electric-field effect on a domain structure in the demagnetized state of MgO/CoFeB with a perpendicular magnetic easy axis. We observe the domain structure with a spatially isotropic maze pattern, whose period is a function of the magnitude of the applied electric field. The electric-field dependence of the domain period indicates that there is electric-field modulation of the exchange stiffness constant of $1.3 \times 10^{-21}$ J/V.

This work was supported in part R&D Project for ICT Key Technology of MEXT and Grants-in-Aid for Scientific Research from JSPS (No. 26889007) as well as MEXT (No. 26103002).

**Figure captions**

FIG. 1. Magnetic-field angle $\theta_H$ dependence of ferromagnetic resonance fields $H_R$ at three electric fields $E$. Solid lines are fitted lines.

FIG. 2. Domain structures at (a) $E = -0.088$ V/nm, (b) 0, and (c) $+0.088$ V/nm. (d)-(f) Two-dimensional fast Fourier transforms (2D-FFTs) of (a)-(c). (g) Averaged amplitudes of FFTs in (d)-(f).

FIG. 3. Electric-field $E$ dependence of characteristic domain periods $D_P$. Circles are experimental results, solid line is obtained from Eq. (1) by assuming the electric field modulates only the magnetic anisotropy, and dotted line is a linear fit.

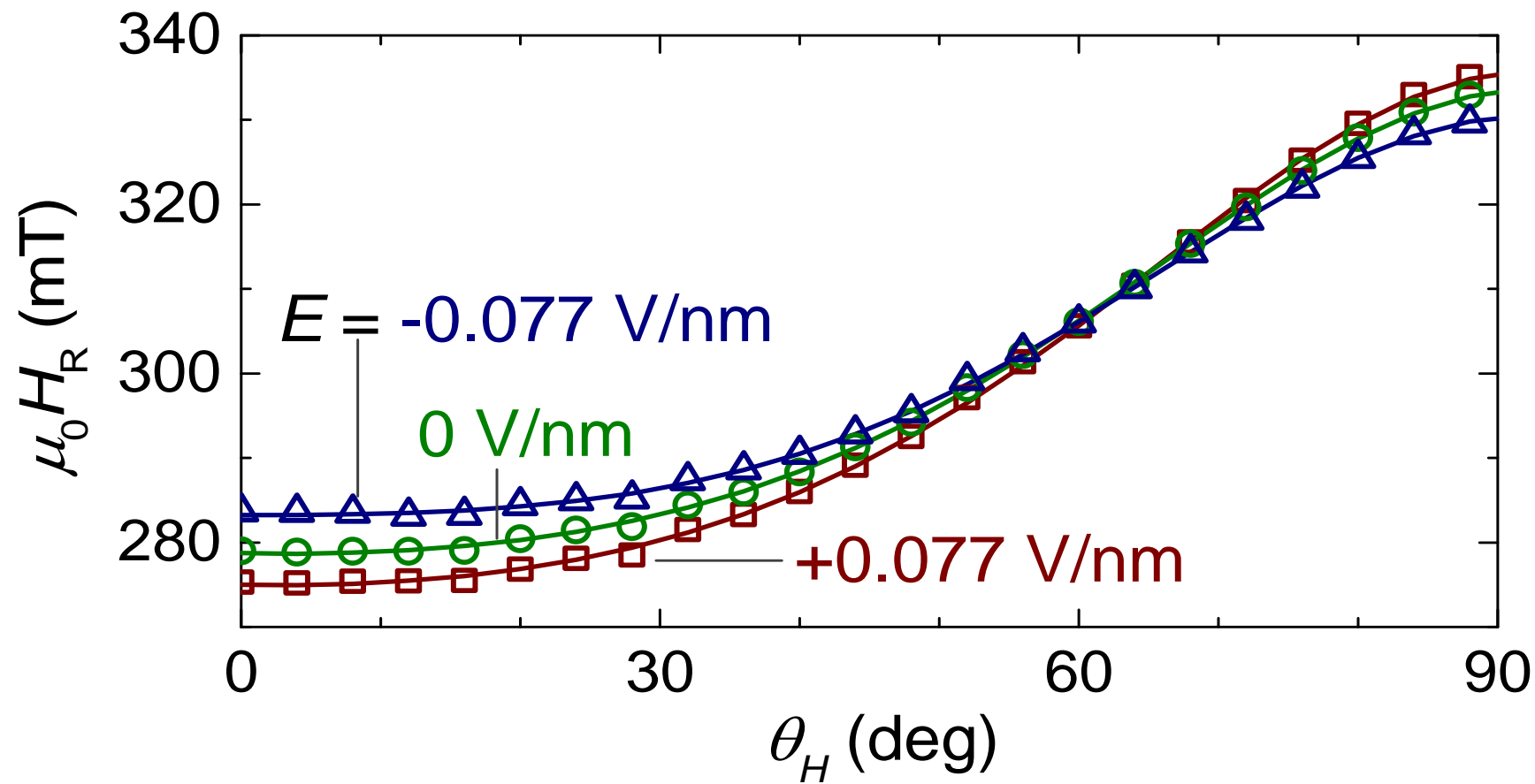

T. Dohi et al., Fig. 1

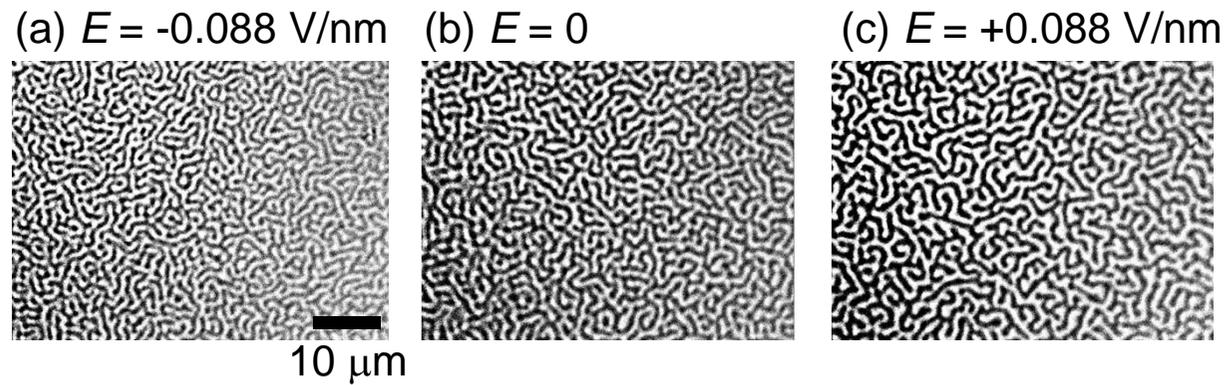
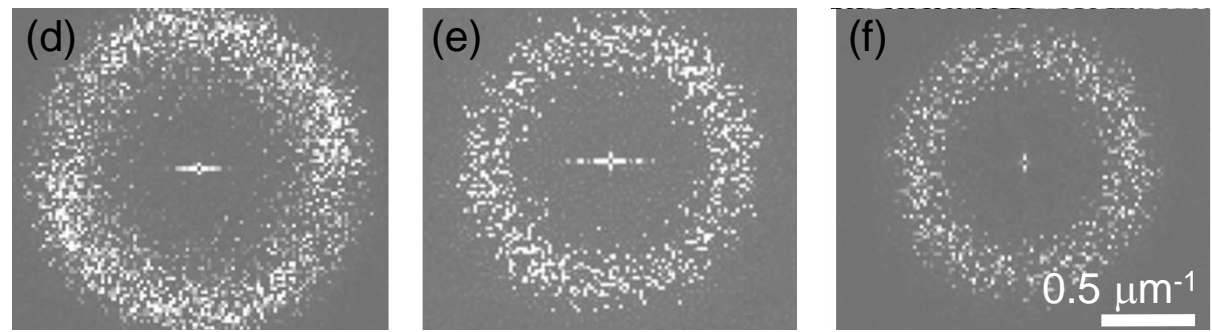
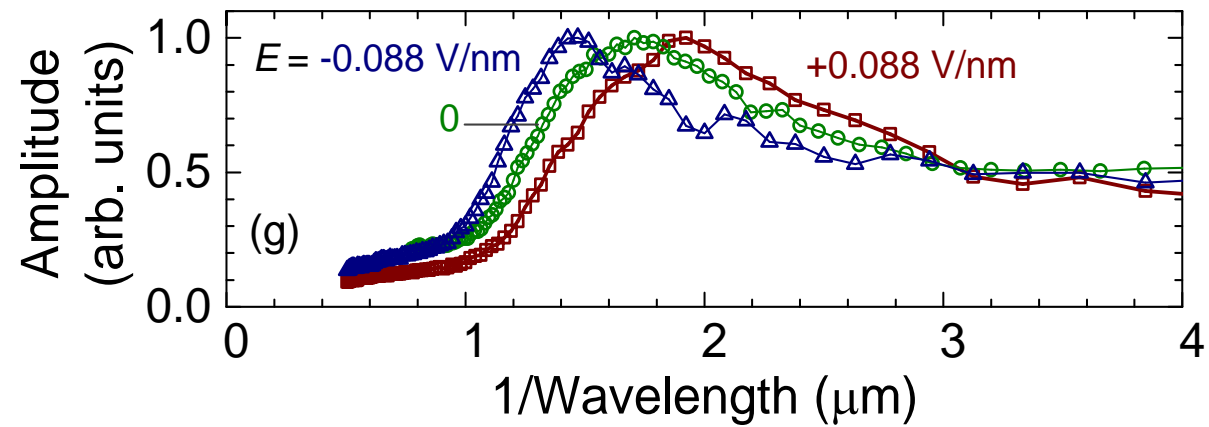

T. Dohi et al., Fig. 3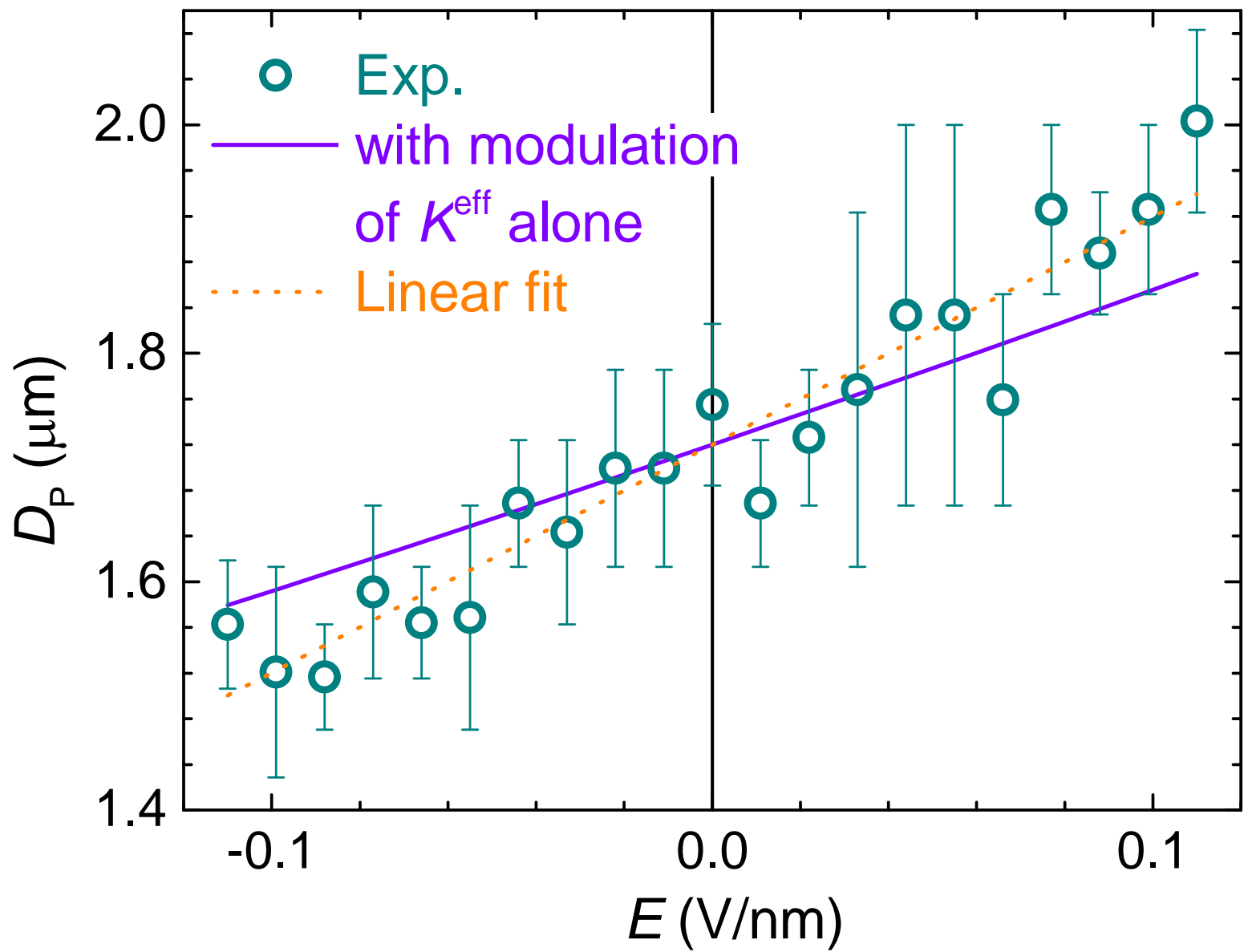